\documentclass[utf8]{frontiersFPHY} 

\usepackage{url,lineno,microtype,subcaption}
\usepackage[onehalfspacing]{setspace}
\setcitestyle{square,authoryear} 

\def\keyFont{\fontsize{8}{11}\helveticabold }
\def\firstAuthorLast{Marsden {et~al.}} 
\def\Authors{Christopher Marsden\,$^{1,*}$, Francesco Shankar\,$^{1}$, Michele Ginolfi\,$^{2}$, Kastytis Zubovas\,$^{3}$}
\def\Address{$^{1}$ Department of Physics and Astronomy, University of Southampton, Highfield, SO17 1BJ, UK \\ $^{2}$ Observatoire de Gen\`{e}ve, Universit\'{e} de Gen\`{e}ve, 51 Ch. des
Maillettes, 1290 Versoix, Switzerland \\ $^{3}$ Center for Physical Sciences and Technology, Sauletekio al. 3, Vilnius LT-10257, Lithuania}
\def\corrAuthor{Christopher Marsden}

\def\corrEmail{c.marsden@soton.ac.uk}

\begin{document}
\onecolumn
\firstpage{1}

\title[The fundamental $M_{\rm BH}$-$\sigma$]{The case for the fundamental $M_{\rm BH}$-$\sigma$ relation}

\author[\firstAuthorLast ]{\Authors} 
\address{} 
\correspondance{} 

\extraAuth{}

\maketitle

\begin{abstract}

\section{}Strong scaling relations between host galaxy properties (such as stellar mass, bulge mass, luminosity, effective radius etc) and their nuclear supermassive black hole's mass point towards a close co-evolution. In this work, we first review previous efforts supporting the fundamental importance of the relation between supermassive black hole mass and stellar velocity dispersion ($M_{\rm BH}$-$\sigma_{\rm e}$). We then present further original work supporting this claim via analysis of residuals and principal component analysis applied to some among the latest compilations of local galaxy samples with dynamically measured supermassive black hole masses. We conclude with a review of the main physical scenarios in favour of the existence of a $M_{\rm BH}$-$\sigma_{\rm e}$ relation, with a focus on momentum-driven outflows.

\tiny
 \keyFont{ \section{Keywords:} Supermassive Black Holes, Velocity Dispersion, Galaxies, Scaling Relations, Principal Component Analysis} 
\end{abstract}

\section{Introduction}

Observational evidence suggests that most local galaxies host a central supermassive black hole (henceforth simply `black hole', not to be confused with an `ordinary' stellar mass black hole). Indeed, galaxies for which high-resolution data can be acquired show stellar kinematic patterns strongly suggesting the presence of a central massive dark object \citep{2005SSRv..116..523F, 2013ARA&A..51..511K}. The central black hole masses, inferred from dynamical measurements of the motions of stars and/or gas in the host galaxies, appear to scale with galaxy-wide properties (or perhaps \textit{bulge}-wide properties), such as stellar mass \citep{1998AJ....115.2285M, 2004ApJ...604L..89H} and stellar velocity dispersion \citep{2000ApJ...539L..13G, 2002ApJ...578...90F, 2002ApJ...574..740T, 2009ApJ...698..198G, 2013ApJ...764..184M, 2015MNRAS.446.2330S}. The existence of such correlations is remarkable, as the black hole's (sub-parsec scale) sphere of influence is orders of magnitude smaller than the scale of it's host galaxy (kilo-parsec scale). These correlations  suggest a close link (a ``co-evolution'') between black holes and host galaxies \citep{1998A&A...331L...1S, 2004ApJ...600..580G}.

The existence of massive black holes at the centre of galaxies also lends further support to the widely-accepted paradigm that quasars, and more generally Active Galactic Nuclei (AGN), are powered by matter accreting onto a central black hole. The release of gravitational energy from an infalling body of mass $m$ approaching the Schwarzschild radius $R_s=2GM/c^2$ of a compact object of mass $M$, is in fact one of the most efficient known processes to release enough energy to explain the large luminosities in AGN. As discussed by \cite{1997iagn.book.....P}, the emission from release of gravitational energy increases with the compactness of the source $M/r$. Assuming that most of the gravitational energy $E$ powering the emission from an accreting black hole originates from within a few times $R_s$, say $r=5R_s$, one could set $E=GMm/5R_s$, implying $E=0.1mc^2$. The latter efficiency $\eta \sim 0.1$ of energy conversion in units of the rest-mass energy, is orders of magnitude higher than the efficiency in stellar fusion ($\eta \sim 0.008$). Theoretical models have also suggested that the energy/momentum release from the central black hole, routinely known as ``AGN feedback'', could have profound consequences on the fate of its host galaxy, potentially driving out a galaxy's gas reservoir, quenching star formation, and shaping the above-mentioned scaling relations \citep{1998A&A...331L...1S, 2019SAAS...48...95K}.

The most prominent and studied scaling relations relate the black hole mass $M_{\rm BH}$ to the stellar velocity dispersion $\sigma_{\rm e}$ \citep{2000ApJ...539L...9F} and the (stellar) mass of the host bulge, $M_{\rm bulge}$ (and by extension the luminosity of the bulge $L_{\rm bulge}$, see  \cite{2003ApJ...589L..21M}). Other types of correlations have been proposed in the literature, such as correlations with the bulge light concentration $c_{\rm bulge}$ \citep{2001ApJ...563L..11G} and even the mass of the surrounding dark matter halo $M_{\rm halo}$ \citep{2002ApJ...578...90F}. This review will focus on the $M_{\rm BH}$-$\sigma_{\rm e}$ relation, where $\sigma_{\rm e}$ is the stellar velocity dispersion inferred from spectral absorption lines (see \cite{2010gfe..book.....M}, Chapter 2).

The $M_{\rm BH}$-$\sigma_{\rm e}$ relation has attracted the attention of the astronomical community since its discovery \citep{2000ApJ...539L...9F}, as it is believed to be closely connected to the galaxy/halo gravitational potential well, and thus may be related to the above-mentioned AGN feedback process \citep{2004ApJ...600..580G}, as further discussed in Section 3.2. The relation is of the form:

\begin{equation}
    \label{eq:badmsigma}
    \log\frac{M_{\rm BH}}{M_\odot} = \alpha + \beta \log\frac{\sigma_{\rm e}}{200kms^{-1}} \, .
\end{equation}

\cite{2000ApJ...539L...9F} initially retrieved a normalization and slope of, respectively, $\alpha=8.14\pm1.80$ and $\beta=4.80\pm0.54$, whereas more recent work \cite[e.g.,][]{2007ApJ...663...53T} suggests $\alpha = 8.21 \pm 0.06$ and $\beta = 3.83 \pm 0.21$. There is some debate in the literature concerning the exact shape of the $M_{\rm BH}$-$\sigma_{\rm e}$ and its dependence on, for example, morphological type or even environment (see, e.g., \cite{2007ApJ...662..808L}, \cite{2006MNRAS.365.1082W} and \cite{2008MNRAS.386.2242H} for more details). It has been noted (e.g. \cite{2012Natur.491..729V}) that several overmassive black holes exist on this relation, hosted by galaxies that have undergone fewer than usual mergers, in tension with semi-analytic models \citep{2015MNRAS.446.2330S}. However, these outliers could simply be the result of incorrect modelling of the galactic bulge/disc \citep{2016MNRAS.457..320S}.

Several groups have noted that the $M_{\rm BH}$-$\sigma_{\rm e}$ relation only weakly evolves with redshift (if at all) \citep[e.g.][]{2009arXiv0908.0328G, 2013ApJ...764...80S, 2015ApJ...805...96S}. Supporting work by other groups base their conclusions on direct estimates of the $M_{\rm BH}$-$\sigma_{\rm e}$ relation on high redshift quasar samples \citep{Woo_2008}, and studies based on comparing the cumulative accretion from AGN with the local black hole mass density, retrieved from assigning to all local galaxies a black hole mass via the $M_{\rm BH}$-$\sigma_{\rm e}$ relation (e.g. \cite{2009ApJ...694..867S}, \cite{2012ApJ...761....5Z}).

On the assumption that all local galaxies host a central black hole, scaling relations could in principle allow us to assign black hole masses to all local galaxies without a direct dynamical mass measurements, thus generating large-scale black hole mass statistical distributions such as black hole mass functions or correlation functions (see \cite{2009NewAR..53...57S} and \cite{2015ApJ...798...54G} for more focused reviews on this topic). For example, a number of groups have used luminosity, as performed by \cite{2004MNRAS.354.1020S}, \cite{1999MNRAS.307..637S} and \cite{2004MNRAS.351..169M}, or even Sersic index, as performed by \cite{2007MNRAS.378..198G}), to generate black hole mass functions. This procedure of course relies on two assumptions: firstly, that the observer has correctly identified the surrogate observable of black hole mass, and secondly that the established scaling relation is reliable. For example, the $M_{\rm BH}$-$\sigma_{\rm e}$ and $M_{\rm BH}$-$L_{\rm bulge}$, probably the most commonly used relations, have led to different black hole mass function estimates (\cite{2007ApJ...670..249L}, \cite{2007ApJ...663...53T}).

An important question is whether the same black hole-galaxy scaling relations hold for both active and inactive galaxies. Several groups suggest that this is indeed the case \citep[e.g.][]{2015ApJ...813...82R, caglar2019llama, 2019MNRAS.485.1278S}. It is important to stress that the samples of nearby (inactive) galaxies on which the black hole-host galaxy relations are based, still remain relatively small, only comprising around $\sim70$-$80$ objects. A key difficulty relies of course in acquiring sufficiently high-resolution data to allow for dynamical black hole mass measurements (see \cite{1999AdSpR..23..925F}, \cite{2005SSRv..116..523F} and \cite{2013ARA&A..51..511K} for reviews of observational challenges).

Indeed, there is a growing body of work (\cite{2007ApJ...660..267B}, \cite{2015ApJS..218...10V}, \cite{2016MNRAS.460.3119S}, \cite{2017MNRAS.466.4029S}, \cite{2019MNRAS.485.1278S}, \cite{2019arXiv191010175S}) supporting the view that that current dynamical black hole mass samples may indeed be ``biased-high'', possibly due to angular resolution selection effects (see \cite{2013degn.book.....M}), with meaningful consequences for any study based on the ``raw'' relations.
Interestingly, \cite{2016MNRAS.460.3119S} showed that, via aimed Monte Carlo simulations, irrespective of the presence of an underlying resolution bias,
the raw and ``de-biased'' scaling relations would still share similar slopes and overall statistical properties (e.g., very similar residuals around the mean), with (noticeable) differences arising only in the  normalization between observed and de-biased scaling relations. In particular, the $M_{\rm BH}$-$\sigma_{\rm e}$ was shown to be more robust and the least affected by possible angular resolution effects.

The main aim of this work is threefold: i) to review the evidence in favour of the primary importance of the $M_{\rm BH}$-$\sigma_{\rm e}$ relation above other black hole scaling relations, ii) to provide further support to velocity dispersion as the main host galaxy property driving the connection between black holes an their hosts, and iii) to review the main theoretical scenarios that give rise to the  $M_{\rm BH}$-$\sigma_{\rm e}$ relation, with a focus on momentum-driven outflows. In sections \ref{section:residuals} and \ref{section:pca} we will describe original evidence based on residuals and principle component analysis (respectively) in support of the primary role played by $M_{\rm BH}$-$\sigma_{\rm e}$. In section \ref{section:theory} we include a description of the theoretical scenarios behind the physical origin of the  $M_{\rm BH}$-$\sigma_{\rm e}$ relation. We then conclude in section \ref{section:conclusion}.

Where cosmological parameters are required, we set $h=0.7$, $\Omega_m=0.3$ and $\Omega_\Lambda=0.7$.

\section{The case for Velocity dispersion}

\label{section:VD}

\subsection{Review of previous work}

Standard regression analyses showed that the $M_{\rm BH}$-$\sigma_{\rm e}$ has the lowest \textit{intrinsic scatter} of any black hole scaling relation; e.g. \cite{2009ApJ...698..198G}, \cite{2016ApJ...818...47S} and \cite{2016ApJ...831..134V}. This alone suggests $\sigma_{\rm e}$ is different from other variables. \cite{2012MNRAS.419.2497B} came to the conclusion that $M_{\rm BH}$ was fundamentally driven by $\sigma_{\rm e}$ due to its relative tightness. This work also tested the possibility for multi-dimensional relations, concluding that the introduction of additional variables barely reduced the scatter with respect to the $M_{\rm BH}$-$\sigma_{\rm e}$, suggesting that stellar velocity dispersion remains a fundamental driving parameter. The amount of scatter characterizing diverse black hole scaling relations has been studied by several groups (\cite{2003ApJ...589L..21M}, \cite{2007ApJ...669...67H}). \cite{2003ApJ...589L..21M} and \cite{2007ApJ...669...67H} explored the addition of the effective radius $R_{\rm e}$ to $\sigma_{\rm e}$ to create a ``fundamental plane'' in the black hole scaling relations, further discussing in \cite{2007ApJ...669...45H} how this relation naturally arises in their simulations, as a (tilted) correlation between black hole mass and bulge binding energy. This conclusion was supported by \cite{2016ApJ...818...47S}, who argued for a multidimensional relation deriving from the bulge kinetic energy ($M_{\rm bulge}\sigma_{\rm e}^2$), as originally suggested by \cite{2005IJMPD..14.1861F}.

\cite{2019MNRAS.490..600D} presented a systematic study of black hole scaling relations on an improved sample of local black holes, confirming that ``the correlation with the effective velocity dispersion is not significantly improved by higher dimensionality''. The authors concluded that the $M_{\rm BH}$-$\sigma_{\rm e}$ is fundamental over multidimensional alternatives, independent of bulge decompositions. This is in line with \cite{2016ApJ...831..134V}, who claimed that the $M_{\rm BH}$-$M_{\rm bulge}$ is mostly a projection of the $M_{\rm BH}$-$\sigma_{\rm e}$ relation.

On more general grounds it has been suggested that, in terms of galactic scaling relations, velocity dispersion may be statistically more significant and relevant than other galaxy observables (e.g., \cite{2011MNRAS.412..684B}, \cite{2011MNRAS.412L...6B}). \cite{2005AJ....129...61B} analysed the color-magnitude-velocity dispersion relation of a early-type galaxy sample of the Sloan Digital Sky Survey (SDSS), concluding that color-magnitude relations are entirely a consequence of the combination of more fundamental correlations with velocity dispersion.

\cite{2007ApJ...660..267B} noted that the $M_{\rm BH}$-$\sigma_{\rm e}$ and $M_{\rm BH}$-$L_{\rm bulge}$ predict different abundances of black holes, with the former predicting a smaller number of more massive black holes. Interestingly, the combined $\sigma_{\rm e}$-$L$ relation (for the dynamically measured black hole sample, e.g. \cite{2002MNRAS.335..965Y}) is inconsistent with the same relation from the SDSS, with smaller $L_{\rm bulge}$ for given $\sigma_{\rm e}$ (regardless of the band used to estimate luminosity). This suggests that the dynamical sample of local black holes may be biased towards objects with higher velocity dispersion when compared to local galaxies of similar luminosity, which obviously calls into question the accuracy of the raw $M_{\rm BH}$-$\sigma_{\rm e}$ and $M_{\rm BH}$-$L_{\rm bulge}$ relations. While unable to identify the source of the bias, modelling of this effect by \cite{2007ApJ...660..267B} and \cite{2016MNRAS.460.3119S} suggested that the bias in the $M_{\rm BH}$-$\sigma_{\rm e}$ is likely to be small, whereas the $M_{\rm BH}$-$L_{\rm bulge}$ is likely to predict over-massive black holes at a fixed galaxy (total) luminosity/stellar mass.

\subsection{Residuals analysis}
\label{section:residuals}

We start by revisiting the residual analysis on the black hole scaling relations following the method outlined in \cite{2016MNRAS.460.3119S}, \cite{2017MNRAS.466.4029S} and \cite{2019MNRAS.485.1278S}. Residuals in pairwise correlations \citep{2012MNRAS.422.1825S} allow for a statistically sound approach to probe the relative importance among variables in the scaling with black hole mass. Residuals are computed as
\begin{equation}
\Delta(Y|X)\equiv\log Y-\langle \log Y|\log X \rangle \,
\label{eq|resid}
\end{equation}
where the residual is computed in the $Y$ variable (at fixed $X$) from the log-log-linear fit of $Y(X)$ vs $X$, i.e. $\langle \log Y|\log X \rangle$. For each pair of variables, each residual is computed 200 times, and at each iteration five objects at random are removed from the original sample. From the full ensemble of realizations, we then measure the mean slope and its 1$\sigma$ uncertainty.

\begin{figure*}
    \center{\includegraphics[width=\textwidth]{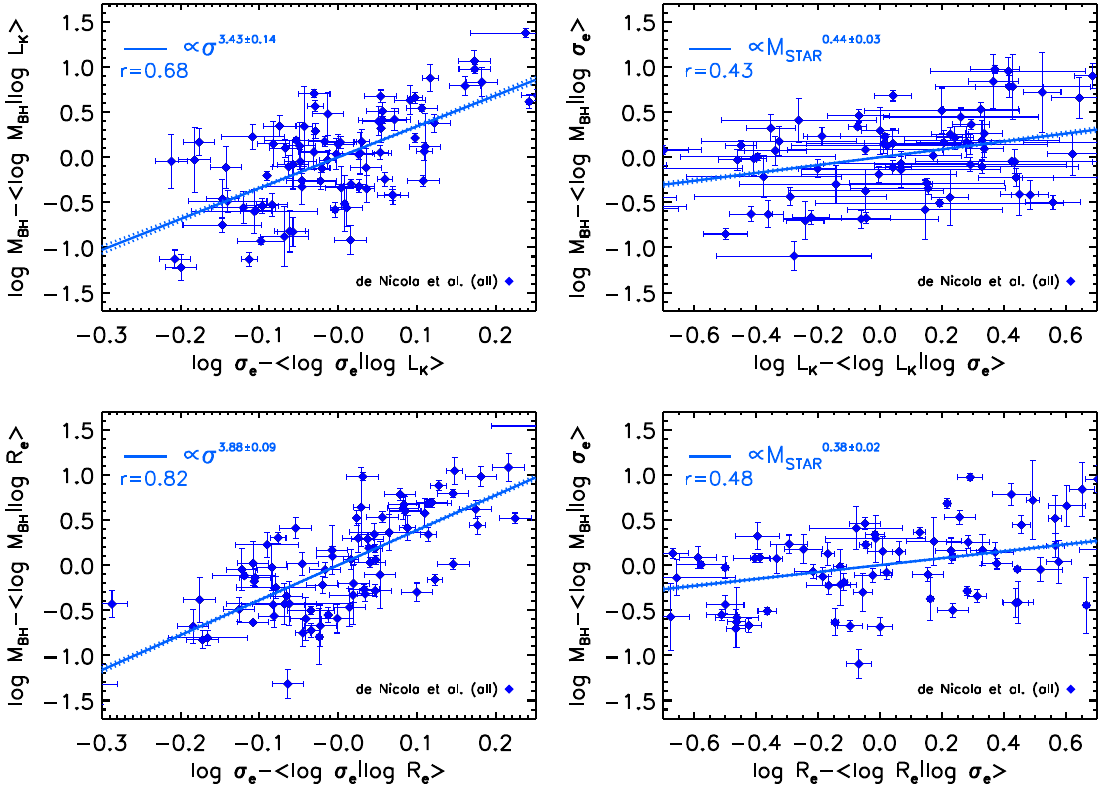}
    \caption{Correlations between residuals from the observed scaling relations, as indicated. The residuals are extracted from the recent homogeneous sample calibrated by \cite{2019MNRAS.490..600D}. It can be clearly seen that black hole mass is strongly correlated with velocity dispersion at fixed galaxy luminosity with a Pearson coefficient $r\sim0.7$ (top left panel), and even more so at fixed effective radius with $r\sim 0.8$ (bottom left panel). Correlations with other relations appear less strong (right panels).
    \label{fig|FigureResiduals}}}
\end{figure*}

\begin{figure*}
    \center{\includegraphics[width=\textwidth]{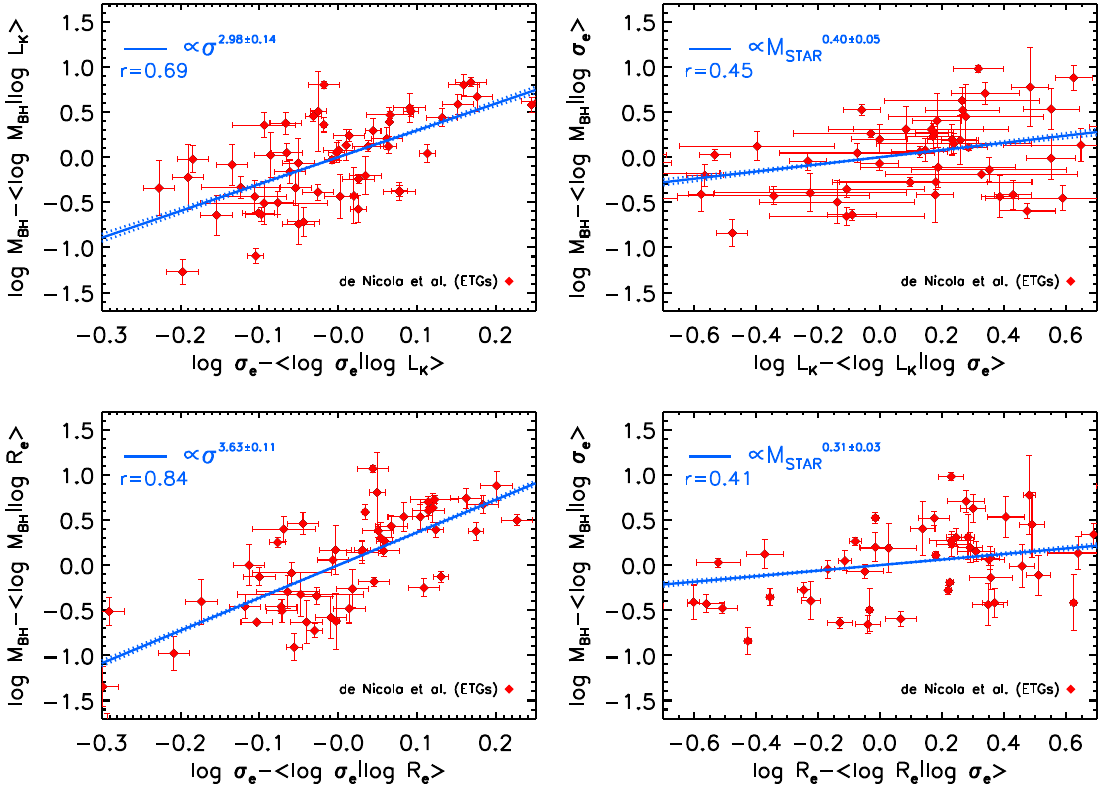}
    \caption{Identical analysis to Figure \ref{fig|FigureResiduals}, but only early type galaxies. Correlations with velocity dispersion are comparable.
    \label{fig|FigureResidualsETGS}}}
\end{figure*}

\begin{figure*}
	\centering
	\includegraphics[width=1\textwidth]{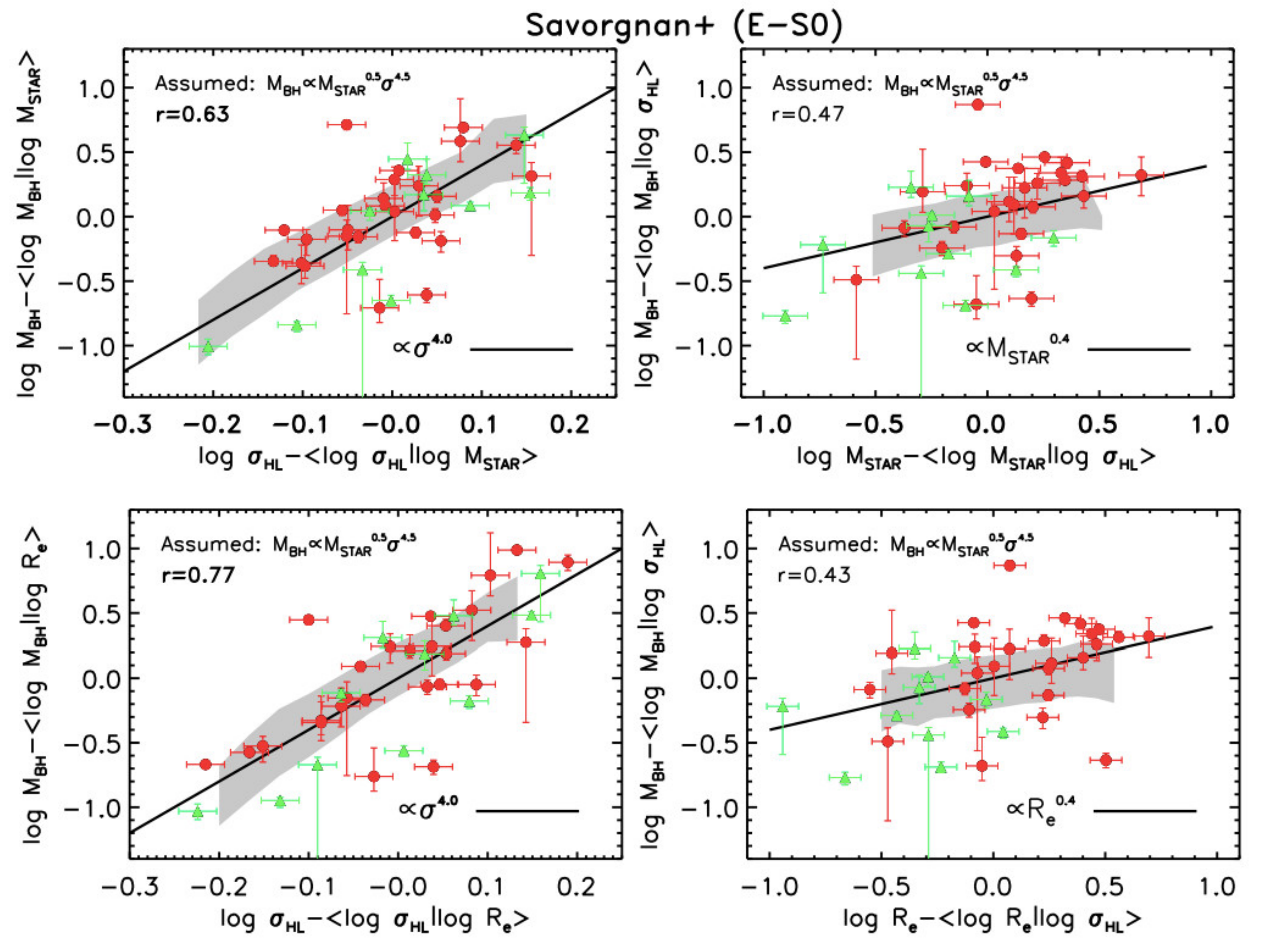}
	\caption{
	Figure 5 from \cite{2016MNRAS.460.3119S} showing correlations between residuals. Correlations with velocity dispersion (left panels) appear to be stronger than other relations. The data is from the sample of \cite{2016ApJ...817...21S}.
	}
	\label{fig:FS16}
\end{figure*}

Our results are shown in Figures \ref{fig|FigureResiduals} and \ref{fig|FigureResidualsETGS}, which show the residuals extracted from the recent homogeneous sample calibrated by \cite{2019MNRAS.490..600D}. Figure \ref{fig|FigureResiduals} shows that black hole mass strongly correlates with velocity dispersion at fixed galaxy luminosity with a Pearson coefficient $r\sim0.7$ (top left panel), and even more so at fixed effective radius with $r\sim 0.8$ (bottom left panel), while the corresponding correlations with stellar luminosity or effective radius are significantly less strong with $r\sim0.4$ at fixed velocity dispersion (right panels). Figure \ref{fig|FigureResidualsETGS} shows the residuals restricting the analysis to only early type galaxies (red circles). The residuals appear quite similar in both slopes and related Pearson coefficients. These results further support the findings by \cite{2016MNRAS.460.3119S} (shown, for comparison, in figure \ref{fig:FS16}) that velocity dispersion is more fundamental than effective radius and stellar mass, and that even disc-dominated galaxies follow similar scaling relations.

The total slope of the $M_{\rm BH}$-$\sigma_{\rm e}$ relation can be estimated as $M_{\rm BH}\propto\sigma^{\beta}M_*^{\alpha}\propto\sigma^{\beta+\alpha\,\gamma}$, where $\gamma$ comes from $M_*\propto\sigma^{\gamma}$. Since SDSS
galaxies tend towards $\gamma\approx 2.2$ \citep{2017MNRAS.466.4029S}, and the residuals in Figure \ref{fig|FigureResiduals} yield $\beta\sim 3$ and $\alpha\sim 0.4$, one obtains a total dependence of $M_{\rm BH}\propto\sigma_{\rm e}^{5}$, consistent with models of black hole growth being regulated by AGN feedback, as further discussed in Section 3.2 (e.g. \cite{1998A&A...331L...1S}, \cite{1999MNRAS.308L..39F}, \cite{2003ApJ...596L..27K}, \cite{2004ApJ...600..580G}).

\subsection{PCA analysis}
\label{section:pca}

We will now present additional \textit{original} work in favour of the $M_{\rm BH}$-$\sigma_{\rm e}$ being more fundamental, via Principal Component Analysis (PCA; \citealp{Jolliffe1986}), which is a powerful complementary statistical technique to the residuals analysis presented above. PCA is a mathematical procedure that diagonalises the covariance matrix of variables in a dataset, providing a set of uncorrelated linearly transformed parameters, called principal components, defined by a set of orthogonal eigenvectors.
The new orientation ensures that the first principal component (PC1) contains as much as possible of the variance in the data, and that the maximum of the remaining variance is contained in each succeeding orthogonal principal component (PC2, PC3, etc.).
In other words, PCA finds the optimal projection of a number of (possibly correlated) physical observables into a smaller number of uncorrelated variables, revealing which quantities are more responsible for the variance (or, in some sense, for the information) in the dataset. PCA has been widely adopted in extragalactic astronomy, for instance to search for possible \textit{dimensionality reduction} of the parameter space necessary to describe a sample (e.g., \citealp{LaraLopez2010, Hunt2012})
or to study the mutual dependencies between observed gas- and metallicity-based galaxy scaling relations (e.g., \citealp{Bothwell2016, Hunt2016, Ginolfi2019}).
Here we use PCA as an alternative technique to explore the black hole scaling relations.
In detail, by quantifying through PCA the robustness of the correlations between $M_{\rm BH}$ and, in turn, $\sigma_{\rm e}$, $L$ (total luminosity) and $R_{\rm e}$ (the bulge effective radius), we can infer which of these observables provides a more fundamental scaling relation.

\subsubsection{Black Hole scaling relations}\label{sec:2DPCA}

In the PCA analysis we continue to use the dataset from \cite{2019MNRAS.490..600D}. To ensure that quantities with a higher dispersion are not over-weighted, we normalize our variables to their mean values, dividing by the standard deviation of their distributions. We therefore define the new variables (for convenience, in what follows we simply define $L=L_K$):
\begin{equation}
\log(M_{\rm BH})^{\rm PCA} = [\log(M_{\rm BH}) - 8.43]/0.99
\end{equation}
\begin{equation}
\log(\sigma_{\rm e})^{\rm PCA} = [\log(\sigma_{\rm e}) - 2.30]/0.18
\end{equation}
\begin{equation}
\log(L)^{\rm PCA} = [\log(L) - 10.92] /0.75
\end{equation}
\begin{equation}
\log(R_{\rm e})^{\rm PCA} = [\log(R_{\rm e}) - 0.34]/0.69
\end{equation}
We perform three different PCA on the 2D-space datasets formed by $M_{\rm BH}$ and, in turn, one among $\sigma_{\rm e}$, $L$ and $R_{\rm e}$. The resulting principal component coefficients are reported in Table 1.
We account for uncertainties in our results following a commonly adopted method (see e.g., \citealp{Bothwell2016, Ginolfi2019}).
We perform a Monte Carlo bootstrap running $10^5$ iterations, in each of which we perturb all the analysed quantities by an amount randomly extracted in a range of values defined by their respective measurement errors.
Thus, the reported principal component's coefficients and their errors are computed, respectively, from the average and the standard deviation of the values obtained over all the iterations.
\newline

\begin{figure*}
	\centering
	\includegraphics[width=1\textwidth]{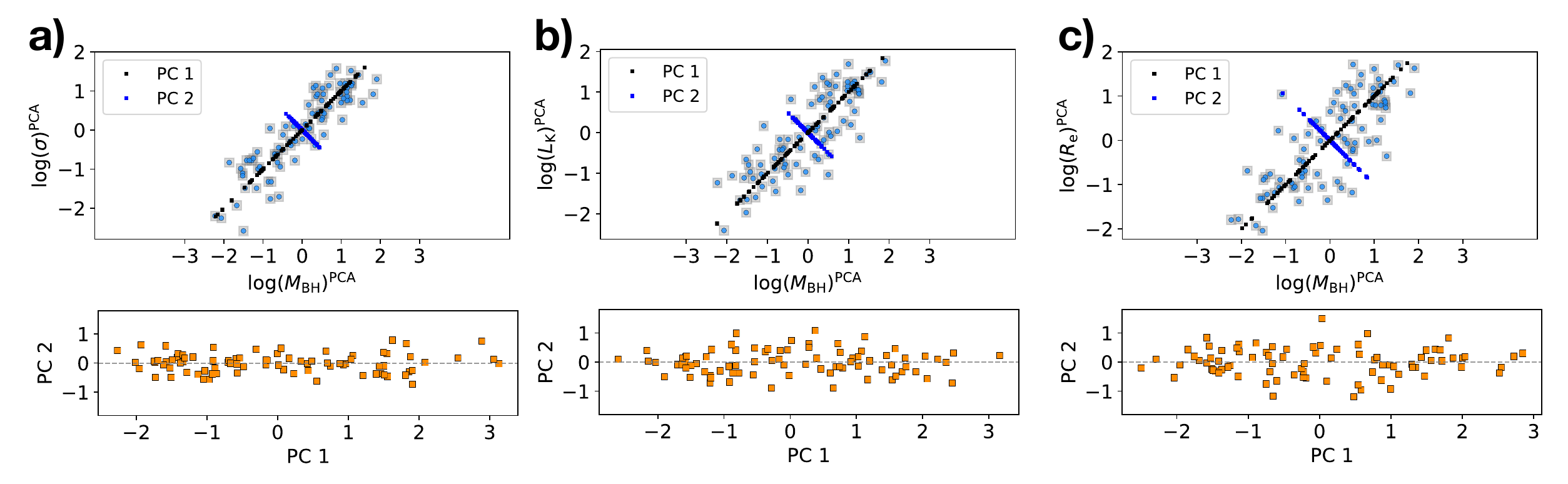}
	\caption{
		Upper panels: the orientations of the mutually orthogonal eigenvectors resulting from our 2D-space PCA are drawn onto the  log($M_{\rm BH}$)$^{\rm PCA}$-log($\sigma_{\rm e}$)$^{\rm PCA}$ (a), log($M_{\rm BH}$)$^{\rm PCA}$-log($L_{\rm k}$)$^{\rm PCA}$ (b), and log($M_{\rm BH}$)$^{\rm PCA}$-log($R_{\rm e}$)$^{\rm PCA}$ (c) planes.
		Lower panels: the projections of the three 2D-space datasets into the principal components are shown.
	}
	\label{fig:fig_PCA_1}
\end{figure*}

In the upper panels of Figure \ref{fig:fig_PCA_1} we show the determined mutually orthogonal eigenvectors drawn onto the planes defined by the 2D-space datasets consisting of log($M_{\rm BH}$)$^{\rm PCA}$ and, in turn, log($\sigma_{\rm e}$)$^{\rm PCA}$, log($L$)$^{\rm PCA}$ and log($R_{\rm e}$)$^{\rm PCA}$.
The three datasets, projected into the principal components, are shown in the lower panels of Figure \ref{fig:fig_PCA_1}.
We find that, although in all three cases PC2 contains only a small fraction of the total variance (see Table 1), confirming that an overall good physical correlation exists among the variables, in the $M_{\rm BH}$-$\sigma_{\rm e}$ relation PC2 is minimised and the dataset can be very well described uniquely by the PC1.
In detail, we find that in the $M_{\rm BH}$-$\sigma_{\rm e}$ relation 95.4 $\pm$ 0.4 \% of the variance is contained into PC1 (with the little remaining information contained in PC2),
while lower amount of variance are contained in the PC1 of the $M_{\rm BH}$ - $L$ relation (89.7 $\pm$ 0.5 \%)
and in the PC1 of the $M_{\rm BH}$ - $R_{\rm e}$ relation (88.2 $\pm$ 0.3 \%).
\newline

Since in all the three cases PC2 contains a little variance, we can set it to zero to obtain a linear approximation of the correlation among our observables from the PCA projected datasets.
Thus, we obtain the following PCA model predictions:
\begin{equation}
\log(M_{\rm BH})^{\rm model}_{\rm \sigma} =  5.4 (\pm0.1) \log(\sigma_{\rm e}) - 4.01 (\pm0.09)
\end{equation}
\begin{equation}
\log(M_{\rm BH})^{\rm model}_{\rm L} =  1.32 (\pm0.04) \log(L) - 6.0  (\pm0.1)
\end{equation}
\begin{equation}
\log(M_{\rm BH})^{\rm model}_{\rm R_{\rm e}} =  1.42 (\pm0.07) \log(R_{\rm e}) + 7.9  (\pm0.1),
\end{equation}

where the non-normalized variables are restored (as defined in the equations discussed above), and the errors on the parameters are computed propagating the uncertainties on the principal component coefficients and on the mean values of the distributions.

\begin{figure*}
	\centering
	\includegraphics[width=1\textwidth]{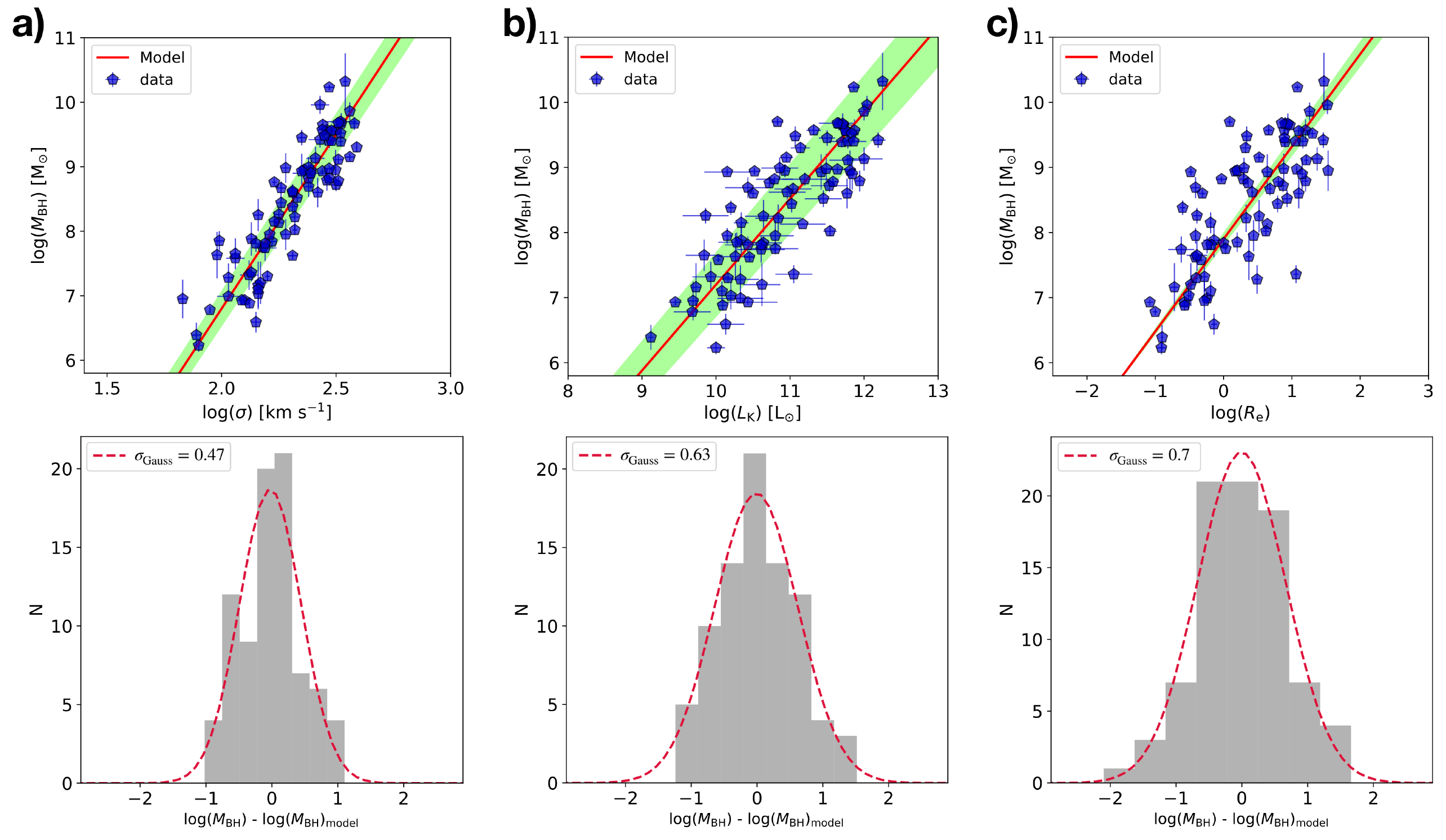}
	\caption{
Upper panels: a  comparison between observations (blue points) and the 2D-space PCA model predictions (red lines) of the $M_{\rm BH}$ - $\sigma_{\rm e}$ (a), $M_{\rm BH}$ - $L_{\rm k}$ (b) and $M_{\rm BH}$ - $R_{\rm e}$ (c) relations is shown. The green shaded regions represent the scatters on the model relations.
Lower panels: the distributions of the residuals are shown (grey histograms), along with their Gaussian fits (red dashed lines).
	}
	\label{fig:fig_PCA_2}
\end{figure*}

In the upper panels of Figure \ref{fig:fig_PCA_2} we show a comparison between the observations in our 2D-space datasets and the PCA model relations, while in the lower panels we show the distributions of the corresponding residuals.
We find that the relation for which our PCA model can better reproduce the data is the $M_{\rm BH}$ - $\sigma_{\rm e}$, with a \textit{Gaussian} 1$\sigma$ scatter of $\sigma \sim 0.47$.
For the $M_{\rm BH}$ - $L$ and $M_{\rm BH}$ - $R_{\rm e}$ relations, our PCA model yields larger scatters in the residuals, respectively $\sigma \sim 0.63$ and $\sigma \sim 0.7$.
These larger scatters are linked to the lower variance contained by PC1 in the samples respect to the $M_{\rm BH}$ - $\sigma_{\rm e}$ case, and therefore a more significant loss of information when setting PC2 to zero.
\newline

Altogether, our 2D-space PCA analysis suggests that, among $\sigma_{\rm e}$, $L$ and $R_{\rm e}$, $\sigma_{\rm e}$ is the observable that better correlates with $M_{\rm BH}$.
The $M_{\rm BH}$-$\sigma_{\rm e}$ is the \textit{more fundamental} scaling relation, with more than 95\% of the information contained in the PC1 and only a scatter of $\sigma \lesssim 0.5$ in the residuals between the data and the PCA model relation.

\subsubsection{4D-space PCA}\label{sec:4DPCA}

As a complementary way of exploring the mutual dependencies among the observables in our sample, we perform a PCA in the 4D-space defined by $M_{\rm BH}$, $\sigma_{\rm e}$, $L$ and $R_{\rm e}$.

We find that PC1 contains 86.1 $\pm$ 0.4 \% of the variance, confirming that the full set of four observables can be approximately well described by a 2D surface.
PC2 contains 9.8 $\pm$ 0.4 \%  of the variance, meaning that accounting for a third dimension could recover $\sim$10 \% of the information, while PC3 and PC4 contains only $\sim 2 ~\%$ (see Table 2).
Following the same scheme discussed in Sec. \ref{sec:2DPCA}, setting to zero the PC that contain less variance, we obtain the best PCA model relation that expresses $M_{\rm BH}$ in terms of the other observables in the dataset:

\begin{equation}
\log(M_{\rm BH})^{\rm model}_{\rm 4D} =  4.05 \log(\sigma_{\rm e}) + 0.64 \log(L) - 0.32 \log(R_{\rm e}) - 7.66.
\end{equation}

\begin{figure}
	\centering
	\includegraphics[width=0.75\columnwidth]{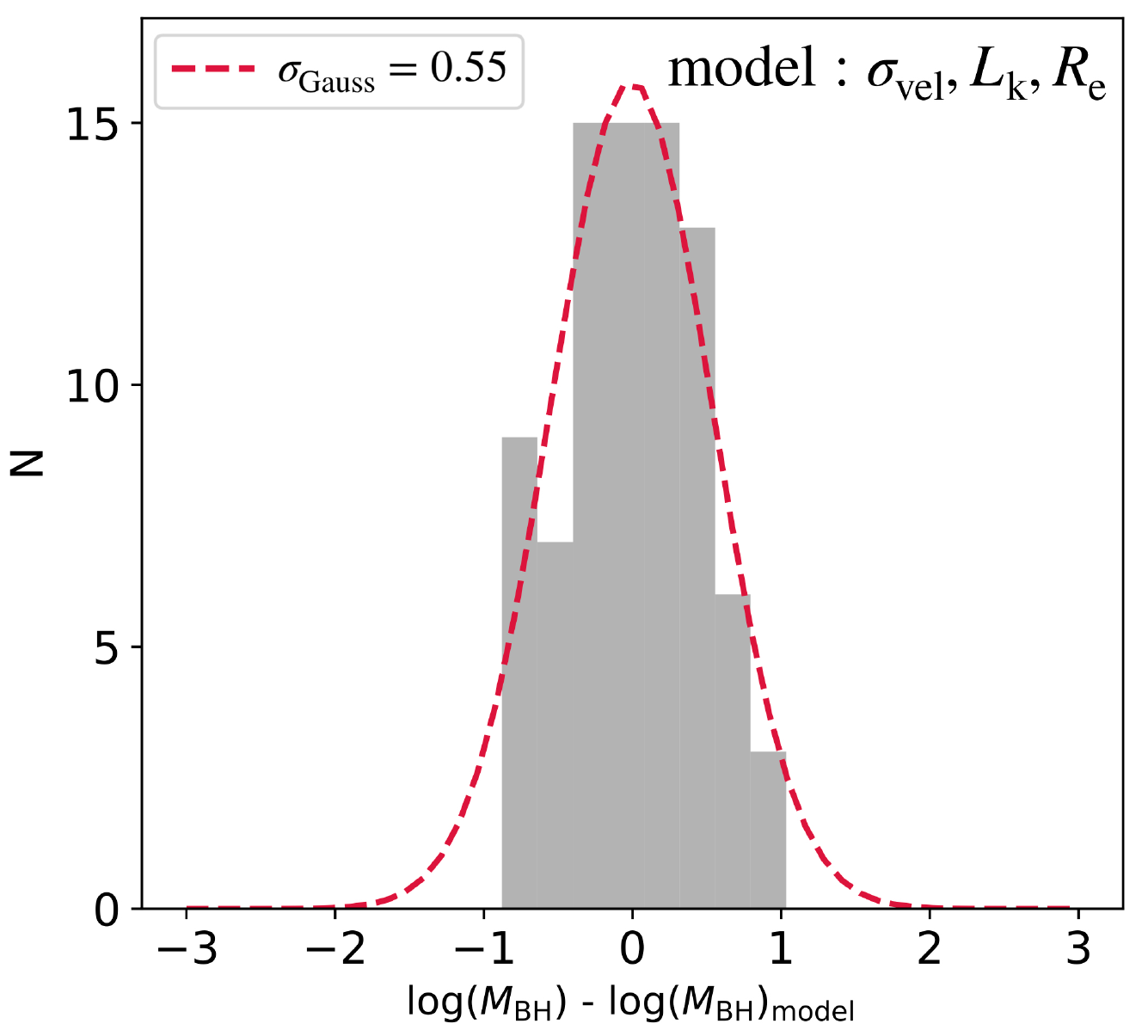}
	\caption{
	The distribution of the residuals computed subtracting the 4D-space PCA model predictions to the observed $M_{\rm BH}$ is shown (grey histogram), along with its Gaussian fits (red dashed line).
	}
	\label{fig:fig_PCA_3}
\end{figure}

Consistently with the result obtained in Section \ref{sec:2DPCA}, we find that the primary dependence is attributed to $\sigma_{\rm e}$, i.e., the quantity that better describes $M_{\rm BH}$.
$L$ and $R_{\rm e}$ have a secondary and tertiary dependence respectively, with relatively much lower weights ($\sim 16\%$ and $\sim 10\%$, computed as the ratios between the coefficients) with respect to $\sigma_{\rm e}$
Interestingly, as shown in Figure \ref{fig:fig_PCA_3}, the residuals obtained from the 4D-space PCA model relation are worse ($\sigma \sim 0.55$) than in the 2D-space PCA model relation obtained trough the optimal projection of the $M_{\rm BH}$-$\sigma_{\rm e}$ space.
This effect is likely to be ascribed to some intrinsic noise introduced when adding $L_{\rm k}$ and $R_{\rm e}$ in a 4D-space.

\section{Theoretical Perspective}
\label{section:theory}

As we have seen, a growing body of work is pointing to the fundamental importance of the $M_{\rm BH}$-$\sigma_{\rm e}$. A key perspective that we have so far neglected in regards to black hole scaling relations is that of the \textit{theoretical modeller}, which we will explore in this section.

The parameters of the galaxy that correlate with $M_{\rm BH}$ tell us which physical processes are most important in setting the black hole mass. Each parameter is related to certain physical quantities. For example, velocity dispersion is naturally related to the mass of the galaxy's spheroidal component, and by extension to its gravitational potential. In the simplest case, modelling the bulge as an isothermal density profile, gas density is $\rho \propto \sigma_{\rm e}^2$ and its weight (the product of the gas mass and gravitational acceleration) is $W \propto \sigma_{\rm e}^4$. Therefore, modelling a connection between the upper limit of the black hole mass and the weight of the gas surrounding it may indeed be a good starting point to explaining the correlation. Alternatively, if the SMBH mass were controlled by stellar processes, such as turbulence driven by stellar feedback, we would expect a strong correlation between $M_{\rm BH}$ and stellar mass. Similarly, if the rate of SMBH feeding from large-scale reservoirs were an important constraint, a correlation with the bulge size $R_{\rm e}$ or dynamical timescale $t_{\rm dyn} \simeq R_{\rm e}/\sigma_{\rm e}$ might emerge. The fact that such correlations are not seen suggests that these processes are secondary to the host's gravitational potential.

A very promising group of models that have emerged over the past two decades are those based on AGN feedback \citep{1998A&A...331L...1S, 2004ApJ...600..580G, Harrison2017NatAs, Morganti2017arXiv}. The common argument is that AGN luminosity transfers energy to the surrounding gas and at some point drives it away, quenching further black hole growth. These models are generally capable of explaining not only the $\sigma_{\rm e}$ relation, but also the presence of quasi-relativistic nuclear winds and large-scale massive outflows observed in many active galaxies. Other models that presume either no causal connection between galaxy and black hole growth \citep{Peng2007ApJ, Jahnke2011ApJ} or those that claim the black hole to be merely a passive recipient of a fraction of the gas used to build up the bulge \citep{Haan2009ApJ, Angles2013ApJ, Angles2015ApJ} make no predictions regarding outflows and generally connect the black hole mass to the mass, rather than velocity dispersion, of the galaxy bulge.

There are several ways of transferring AGN power to the surrounding gas, e.g. radiation, winds and/or jets \citep{Morganti2017arXiv}. Jets are typically efficient on galaxy cluster scales, heating intergalactic gas and prevent it from falling back into the galaxy \citep{McNamara2007ARA&A}. This process, referred to as ``maintenance mode'' of feedback, prevents the SMBH mass from growing above the limit established by the $M_{\rm BH}$-$\sigma_{\rm e}$ relation. Jets are considered to be the primary form of feedback in AGN that accrete at low rates and have luminosities $L < 0.01 L_{\rm AGN}$ \citep{Merloni2007MNRAS}. The opposite type of feedback is known as ``quasar mode'', and it is believed to be most efficient in more luminous AGN. Here, again, there are two possibilities in which energy can be transferred. Directly coupling AGN luminosity to the gas in the interstellar medium is possible if the gas is dusty (due to a very high opacity, see \cite{Fabian2008MNRAS}). On the other hand, dust evaporates when shocked to the temperatures expected within AGN outflows \citep{Barnes2018arXiv}, potentially limiting the impact of radiation-driven outflows. A much more promising avenue is to connect the AGN with the surrounding gas via a quasi-relativistic wind \citep{King2015ARA&A}. Such a model naturally produces both a $M_{\rm BH}$-$\sigma_{\rm e}$ relation similar to the observed one, and outflow properties in excellent agreement with observations, both within galaxies \citep{Zubovas2012ApJ, Cicone2014A&A, Menci2019ApJ} and on intergalactic scales in galaxy groups \citep{Lapi2005ApJ}.

\subsection{AGN wind-driven feedback}

AGN are highly variable on essentially all timescales and are known to occasionally reach the Eddington luminosity
\begin{equation}
  L_{\rm Edd} = \frac{4 \pi G M_{\rm BH} c}{\kappa_{\rm e.s.}},
\end{equation}
where $\kappa_{\rm e.s} \simeq 0.346$~cm$^2$~g$^{-1}$ is the electron scattering opacity. Under such circumstances, the geometrically thin accretion disc produces a quasi-spherical wind that self-regulates to an optical depth $\tau \sim 1$ \citep{King2003MNRAS}. Therefore each photon emitted by the AGN will, on average, scatter only once before escaping to infinity, and the wind carries a momentum rate
\begin{equation}
  \dot{M}_{\rm w} v_{\rm w} = \tau \frac{L_{\rm AGN}}{c},
\end{equation}
where $\dot{M}_{\rm w}$ is the wind mass flow rate, $v_{\rm w}$ is the wind velocity and $L_{\rm AGN} \equiv lL_{\rm Edd}$ is the AGN luminosity, where $l$ is the Eddington ratio. By writing $L_{\rm AGN} = \eta \dot{M}_{\rm BH} c^2$, we find the wind velocity to be
\begin{equation}
  v_{\rm w} = \frac{\tau \eta}{\dot{m}}c,
\end{equation}
where $\dot{m} \equiv \dot{M}_{\rm w}/\dot{M}_{\rm BH}$. The value of $\dot{m}$ is highly uncertain, but should not be extremely different from unity. To see this, consider the extreme ends of the possible range of $\dot{M}_{\rm BH}$. If the accretion rate on to the accretion disc is significantly below Eddington, no wind is produced, while if the accretion rate rises above the Eddington limit, the wind moderates the accretion flow. Overall, the highest possible average accretion rate is the dynamical rate:
\begin{equation}
  \dot{M}_{\rm dyn} = f_{\rm g} \frac{\sigma^3}{G} \simeq \frac{64}{\sigma_{\rm200}}\dot{M}_{\rm Edd},
\end{equation}
where $f_{\rm g} \simeq 0.16$ is the cosmological gas fraction and $\sigma \equiv 200 \sigma_{\rm200}$~km~s$^{-1}$ is the velocity dispersion in the galaxy \citep{King2010MNRASb, King2015ARA&A}. In deriving the second equality, we used the $M_{\rm BH} - \sigma$ relation that is derived below, in eq. \ref{eq:mcritical}. Therefore, in most cases, the SMBH feeding rate is not significantly higher than the Eddington rate, unless $M_{\rm BH}$ is well below the observed relation. As a result, we take $\dot{m} \sim 1$ for the rest of this section. This leads to the final expression for the AGN wind velocity
\begin{equation}
  v_{\rm w} \simeq\eta c \simeq 0.1c,
\end{equation}
which is very close to the average velocity in observed winds \citep{Tombesi2010A&A, Tombesi2010ApJ}. The kinetic power of the wind is
\begin{equation}
  \dot{E}_{\rm w} = \frac{\dot{M}_{\rm w} v_{\rm w}^2}{2} \simeq \frac{\eta}{2}L_{\rm AGN} \simeq 0.05 L_{\rm AGN}.
\end{equation}

The wind rapidly reaches the interstellar medium (ISM) surrounding the AGN and shocks against it. The shock is strong, since $v_{\rm w}/\sigma \gg 1$, and the wind heats up to a temperature
\begin{equation}
  T_{\rm sh} = \frac{3m_{\rm p}v_{\rm w}^2}{16 k_{\rm b}} \simeq 10^{10} {\rm K},
\end{equation}
where $m_{\rm p}$ is the proton mass, and  $k_{\rm b}$ is the Boltzmann constant. The most efficient cooling process at this temperature is Inverse Compton (IC) cooling via interaction with AGN photons \citep{King2003ApJ,Faucher2012MNRAS}. Most of the photons interact with electrons in the shocked wind, and a two-temperature plasma develops \citep{Faucher2012MNRAS}. The actual cooling timescale then depends on the timescale for energy equilibration between electrons and protons. As a result, cooling is highly inefficient and the shocked wind can expand as an approximately adiabatic bubble.

The subsequent evolution of the expanding bubble depends on the density structure of the ISM. Most of the energy stored in the hot wind bubble escapes through the low-density channels and creates a large-scale outflow \citep{Zubovas2014MNRAS}. Denser clouds, however, remain and are mainly affected by the direct push of the wind material. These two situations create two kinds of outflow, known as energy-driven and momentum-driven, respectively. The latter kind is responsible for establishing the $M_{\rm BH} - \sigma_{\rm e}$ relation.

\subsection{The predicted relation}

Momentum-driven outflows push against the dense clouds surrounding the black hole. These clouds are the most likely sources of subsequent black hole feeding, therefore their removal quenches further black hole growth for a significant time and establishes the $M_{\rm BH}$-$\sigma_{\rm e}$ relation \citep{King2003ApJ, Murray2005ApJ, King2010MNRASa}. Considering the balance between AGN wind momentum and the weight of the gas $W_{\rm gas}$ leads to a critical AGN luminosity required for clearing the dense gas:
\begin{equation}
  L_{\rm crit} = W_{\rm gas}c \simeq \frac{4f_{\rm g}\sigma^4c}{G},
\end{equation}
where the second equality assumes that the gas distribution and the background gravitational potential are isothermal, i.e. $\rho = \sigma^2/\left(2\pi G R^2\right)$ \citep{Murray2005ApJ}. Equating this critical luminosity with the Eddington luminosity of the black hole allows us to derive a critical mass \citep{King2010MNRASa}:
\begin{equation}\label{eq:mcritical}
  M_{\rm crit} \simeq \frac{f_{\rm g}\kappa_{\rm e.s.}\sigma^4}{\pi G^2} \simeq 3.2\times10^8 \frac{f_{\rm g}}{0.16} \sigma_{\rm200}^4 \, M_\odot.
\end{equation}
This value is very close to the observed one, although it has a slightly shallower slope. This discrepancy may be explained by the fact that the black hole still grows during the time while it drives the gas away \citep{Zubovas2012MNRASb}. As the gas is pushed away, it joins the energy-driven outflow. This outflow coasts for approximately an order of magnitude longer than the AGN phase inflating it and stalls at a distance \citep{King2011MNRAS}
\begin{equation}
  R_{\rm stall} \simeq \frac{v_{\rm e}^2}{\sigma} t_{\rm AGN},
\end{equation}
where $t_{\rm AGN}$ is the duration of the driving phase and the energy-driven outflow velocity is \citep{King2005ApJ, Zubovas2012ApJ}
\begin{equation}
  v_{\rm e} = \left(\frac{2\eta c}{3 \sigma}\frac{0.16}{f_{\rm g}}\right)^{1/3} \simeq 925 \sigma_{\rm200}^{2/3} \left(\frac{0.16}{f_{\rm g}}\right)^{1/3} {\rm km s}^{-1}.
\end{equation}
By equating $R_{\rm stall}$ with either the bulge radius or the virial radius of the galaxy, we obtain the time $t_{\rm AGN}$ for which the galaxy must be active in order to quench further accretion on to the black hole and find $t_{\rm AGN} \propto R \sigma^{-1/3} \propto \sigma^{2/3}$, since $R \propto \sigma_{\rm e}$ on average \citep[this relation arises from the Fundamental plane of galaxies, see][]{Djorgovski1987ApJ, Cappellari2013MNRAS}. Note that this growth does not need to happen all at once: as long as the outflow is still progressing by the time the next episode begins, the system behaves as if it was powered by a continuously shining AGN \citep{Zubovas2019MNRASa}.

This extra growth steepens the $M_{\rm BH}$-$\sigma_{\rm e}$ relation beyond the simpler analytical prediction and brings it more in line with observations \citep{Zubovas2012MNRASb}. Furthermore, it shows that galaxy radius may be an important secondary parameter determining the final black hole mass.

As a final note, the extra black hole growth while clearing the galaxy also depends on its spin. Since a rapidly spinning black hole produces more luminosity and drives a faster outflow than a slow-spinning one, the latter has to be active for longer and grow more before it clears the gas from the galaxy. Although present-day estimates of black hole spins are not robust or numerous enough to test this prediction in detail, this might become possible in the near future \citep{Zubovas2019MNRASc}.

In general, theoretical models based on momentum-driven outflows are capable of naturally explaining the relationship between black hole mass and velocity dispersion, primarily due to the latter acting as a tracer of the host's gravitational potential well. In addition, these models could account for \textit{secondary}, weaker dependencies on, e.g., galaxy stellar mass or size, which may still be allowed by current data as discussed above (see, e.g., Figures \ref{fig|FigureResiduals} and \ref{fig:fig_PCA_2} and \citealt{2016MNRAS.460.3119S}).

\section{Discussion and Conclusions}
\label{section:conclusion}

In this paper we have reviewed previous evidence for the $M_{\rm BH}$-$\sigma_{\rm e}$ being the most fundamental of all black hole-host galaxy scaling relations (among those discovered so far) and we have presented new evidence based on the statistical analysis of the sample recently compiled by \cite{2019MNRAS.490..600D}. Both residuals (e.g., \cite{2016MNRAS.460.3119S}) and PCA analyses point to $\sigma_{\rm e}$ being more fundamental than both stellar luminosity/mass or effective radius in their correlation to central black hole mass.

Theoretically, as reviewed by \cite{2015ARA&A..53..115K}, the $M_{\rm BH}$-$\sigma_{\rm e}$ arises as a consequence of AGN feedback. In short, the black hole in these models is expected to grow until it becomes massive enough to drive energetic/high-momentum large-scale winds that can potentially remove residual gas, inhibiting further star formation and black hole growth. The limiting mass reached by the black hole, which ultimately depends on the potential well of the host, naturally provides an explanation for the existence of the $M_{\rm BH}$-$\sigma_{\rm e}$ relation.

Its fundamental nature and lower inclination towards selection biases (in comparison to other scaling relations, e.g. \cite{2016MNRAS.460.3119S}) make the $M_{\rm BH}$-$\sigma_{\rm e}$ relation the ideal benchmark for statistical studies of black holes in a variety of contexts. The $M_{\rm BH}$-$\sigma_{\rm e}$ relation should always be the one adopted to constrain the $f_{\rm vir}$ factor used in reverberation mapping studies (see e.g. \cite{2006ApJ...641..689V}) to  infer black hole masses from active galaxies (e.g., \cite{2019MNRAS.485.1278S}). The $M_{\rm BH}$-$\sigma_{\rm e}$ relation also provides more robust large-scale clustering predictions in black hole mock catalogues \citep{2019arXiv191010175S}. Furthermore, pulsar timing array predictions of the gravitational wave background (e.g. \cite{2013CQGra..30v4009K}) are strongly dependent on the normalization of the black hole scaling relations (\cite{2008MNRAS.390..192S}, \cite{2016MNRAS.460.3119S}), but they should be based on the $M_{\rm BH}$-$\sigma_{\rm e}$ rather than on the $M_{\rm BH}$-$M_*$ relation (see \cite{2015MNRAS.451.2417R}).

The shape and scatter of the $M_{\rm BH}$-$\sigma_{\rm e}$ relation could yield important information on the evolutionary channels of black hole growth. For example, its scatter could retain memory of the merger histories of the host galaxies \citep{2015MNRAS.446.2330S}. More broadly speaking, global star formation and black hole growth from continuity equation argument modelling is known to peak at around $z\sim2$ (e.g. \cite{2009ApJ...690...20S}, \cite{2014MNRAS.439.2736D}). This is in itself consistent with the idea that black holes and their hosts may be co-evolving, and understanding how the $M_{\rm BH}$-$\sigma_{\rm e}$ relation precisely evolves over cosmic time or change as a function of environment could set invaluable constraints on the mechanisms behind black hole growth (e.g. \cite{2014MNRAS.442.2304H}, \cite{2015MNRAS.453.4112F} and \cite{2015MNRAS.452..575S}).

\section*{Conflict of Interest Statement}
The authors declare that the research was conducted in the absence of any commercial or financial relationships that could be construed as a potential conflict of interest.

\section*{Author Contributions}

C. Marsden is the primary author and wrote the background and review sections. F. Shankar provided supervision,  edited the manuscript, and provided the calculations of the residuals. M. Ginolfi performed and wrote the section on PCA analysis, K. Zubovas wrote the section on theoretical models.

\section*{Funding}
C. Marsden acknowledges funding from the ESPRC for his PhD. F. Shankar acknowledges partial support from a Leverhulme Trust Research Fellowship.

\section*{Acknowledgments}
We thank Stefano de Nicola and Alessandro Marconi for sharing their data in electronic format. We acknowledge extensive use of the Python libraries astropy, matplotlib, numpy, pandas, and scipy.

\bibliographystyle{frontiersinHLTH&FPHY} 
\bibliography{bib.bib}


\newpage

\section*{Figure captions}


\begin{table}\label{tab:table1}
	\centering
	\begin{tabular}{ll}
		\hline\hline
		~  & ~ \\ [-1.5ex]
		$M_{\rm BH}$ - $\sigma_{\rm vel}$ & Variance [\%]  \\  [+0.5ex]
		~  & ~\\ [-1.5ex]
		PC 1 & 95.4 $\pm$ 0.4  \% \\ [+0.5ex]
		PC 2 & 4.6 $\pm$ 0.3  \  \\ [+0.5ex]
		
		\hline
		
		$M_{\rm BH}$ - $L_{\rm k}$ &  Variance [\%]   \\  [0.5ex]
		~  & ~ \\ [-1.5ex]
		PC 1 & 89.7 $\pm$ 0.5  \  \\ [+0.5ex]
		PC 2 & 10.3 $\pm$ 0.4  \  \\ [+0.5ex]
		
		\hline
		
		$M_{\rm BH}$ - $R_{\rm e}$ &  Variance [\%]\\  [+0.5ex]
		~  & ~ \\ [-1.5ex]
		PC 1 & 88.2 $\pm$ 0.3  \ \\ [+0.5ex]
		PC 2 & 11.8 $\pm$ 0.2  \  \\ [+0.5ex]

		\hline\hline
		
	\end{tabular}
	\caption{The variance percentages contained by the principal components resulting from our PCA on the three 2D-datasets, $M_{\rm BH}$ - $\sigma_{\rm vel}$, $M_{\rm BH}$ - $L_{\rm k}$ and $M_{\rm BH}$ - $R_{\rm e}$, are reported.}
\end{table}

\begin{table*}\label{tab:table2}
	\centering
	\begin{tabular}{llllll}
		\hline\hline
		~  & ~& ~ & ~& ~& ~ \\ [-1.5ex]
		Principal Component & log($\sigma_{\rm vel}$)$^{\rm PCA}$ & log($L_{\rm k}$)$^{\rm PCA}$ & log($R_{\rm e}$)$^{\rm PCA}$ & log($M_{\rm BH}$)$^{\rm PCA}$ & Variance [\%]\\  [+0.5ex]
		~  & ~& ~ & ~& ~& ~  \\ [-1.5ex]
		PC 1 & 0.491 $\pm$ 0.002 		& 0.514	 $\pm$ 0.001      & 0.490 $\pm$ 0.002 		& 0.503 $\pm$ 0.002 & 86.1 $\pm$ 0.4 \\ [+0.5ex]
		PC 2 & 0.14 $\pm$ 0.01 		& -0.09 $\pm$ 0.02       & -0.15 $\pm$ 0.01 	   	& +0.09 $\pm$ 0.02 & 9.8 $\pm$ 0.5\\ [+0.5ex]
		PC 3 & 0.22 $\pm$ 0.04 			  & -0.45 $\pm$ 0.03      	  & +0.38 $\pm$ 0.02 			& -0.13 $\pm$ 0.05 & 2.5 $\pm$ 0.3\\ [+0.5ex]
		PC 4 & 0.39 $\pm$ 0.03 		& +0.25 $\pm$ 0.05      & -0.12 $\pm$ 0.05 		& -0.52 $\pm$ 0.02 & 1.6 $\pm$ 0.3\\ [+0.5ex]
		
		\hline\hline
		
	\end{tabular}
	\caption{The coefficients and the variance of the principal components resulting from our 4D-space ($M_{\rm BH}$-$\sigma_{\rm e}$-$L$-$R_{\rm e}$) PCA are reported.}
\end{table*}

\end{document}